\documentclass[twocolumn,showpacs,amsmath,amssymb,prl]{revtex4}

\usepackage{epsfig}

\begin{document}
\title{Pairing symmetry in the anisotropic Fermi superfluid
under $p$-wave Feshbach Resonance}
\author{Chi-Ho Cheng and Sung-Kit Yip}
\affiliation{ Institute of Physics, Academia Sinica, Taipei,
Taiwan}

\date{\today}

\begin{abstract}
The anisotropic Fermi superfluid of ultra-cold Fermi atoms under
the $p$-wave Feshbach resonance is studied theoretically. The
pairing symmetry of the ground state is determined by the strength of
the atom-atom magnetic dipole interaction. It is $k_z$
for a strong dipole interaction; while it becomes $k_z - i
\beta k_y$, up to a rotation about $\hat z$, for a weak one
(Here $\beta < 1$ is a numerical coefficient).
By changing the external magnetic field
or the atomic gas density, a phase transition between these two
states can be driven. We discuss how the pairing symmetry of
the ground state can be determined in the time-of-flight
experiments.
\end{abstract}


\maketitle


\section{I. Introduction}

The trapped Bose or Fermi atomic gases
\cite{Dalfovo99,Castin01} are usually weakly
interacting since the inter-particle distances between the atoms are
typically
much larger than their scattering lengths.
However, it has been recognized that the
inter-particle interaction can in fact be tuned via Feshbach
resonances \cite{Stwalley76,Tiesinga92}. The interaction changes from weakly
to strongly attractive by varying the external magnetic field
across the Feshbach resonance. For a two-component Fermi gas,
the ground state evolves from a Bardeen-Cooper-Schrieffer (BCS)
 superfluid with
long-ranged (compared with inter-particle distances) Cooper pairing
to a Bose-Einstein condensate (BEC) of tightly bound pairs
\cite{cross0,nozieres,randeria90}.
This cross-over has recently been a subject of
intensive theoretical
\cite{ohashi,falco04,perali,java,numeric}
and experimental
\cite{regal,bourdel,zwierlein,bartenstein,kinast}
investigations.

Most previous investigations deal with $s$-wave Feshbach
resonance. Therefore, both the Cooper pairs and the bound states
between two fermions have $s$-wave symmetries. Recent experiments
demonstrated $p$-wave Feshbach resonances
\cite{Regal03,Ticknor04,Zhang04,Gunter05}. It thus raises the
possibility of $p$-wave Fermi superfluid and BEC of $p$-wave
``molecules" \cite{klinkhamer,ohashi05,ho05} (see also
\cite{alt}).

To begin, we first recall the well-known superfluid
$^3$He \cite{Leggett75}. $^3$He has (nuclear) spin $1/2$, in
general {\it not} spin-polarized and basically spatially
isotropic, that is, the interaction between two atoms
is dependent of the orientation of their relative
distance.
 Back in the 60's, Anderson and Morel \cite{Anderson61}
investigated theoretically the superfluid state for this system by
assuming that the pairing exists only between the same (say
${\uparrow}$) species. They showed that the ground state corresponds to Cooper
pairing in the $l = 1$, $m = 1$ channel, that is, the pairing
wavefunction has the symmetry of the spherical harmonic
$Y_{11}(\hat k) \propto (\hat k_x + i \hat k_y)$
 (or its spatial rotation, such as $\hat k_z - i \hat k_y$).
 This orbital symmetry is
realized in the $^3$He A-phase and is known as the ``axial" state
\cite{Leggett75}. One can thus suspect that for a
spin-polarized but otherwise spatially isotropic system, the
pairing is again in the $Y_{11}(\hat k)$ state.
This has been confirmed in \cite{ho05}.

However, in the atomic gas system, as demonstrated and explained
by Ticknor {\it et al.} \cite{Ticknor04} in the context of
$^{40}{\rm K}$,
 the magnetic dipole interaction, which breaks
rotational symmetry, may be important.
(An Alkali atom has one additional electron beyond
closed electronic shell(s), hence it
possesses a magnetic moment mainly due to
this extra electron, though the precise value of
this moment depends on the hyperfine
interaction and the external magnetic field).
They show that, due to this
magnetic dipole  interaction,
 for magnetic field along $\hat z$, the $l=1$, $m=0$
resonance occurs at a higher magnetic field than the $m = {\pm} 1$
ones (see also \cite{Gunter05}) For a given magnetic field, the
induced effective interaction is anisotropic. In our previous
paper \cite{Cheng05} (see also \cite{gurarie}), we discussed
theoretically the expected ground state order parameter under this
circumstance. Since the interaction of the $m = 0$ channel is more
attractive than that of the $m = {\pm} 1$, in the case that the $m
= 0$ and $m = {\pm} 1$ resonances are sufficiently far apart, we
showed that the pairing can only occur in the $m=0$ channel, and
the ground state becomes $Y_{10}(\hat k) \propto k_z$. The system
is either the BCS state with $k_z$ Cooper pairing or the BEC state
of bosonic ``molecules" in the $m=0$ channel, depending on the
detuning. Fermion pairing or bosons in $m=\pm 1$ channel do not
exist. The orbital symmetry of the pairing is the same as the
``polar" phase in the $^3$He literature \cite{Leggett75}.

For the case that two resonances in $m=0$ and $m=\pm 1$ channels
are closed to each other, we showed  that the ground state
symmetry is $k_z - i \beta k_y$ up to rotation about $\hat z$,
where $\hat z$ is along the magnetic field direction and $\beta <
1$. This state is thus intermediate between the axial $(\beta =
1)$ the the polar $(\beta = 0)$ phases.

This paper provides some details of our earlier investigation \cite{Cheng05},
as well as some additional information.  It
is organized as follows. We first study the Cooper
pairing symmetry of a weak-coupling BCS model,
where the p-wave pairing interaction is anisotropic.
We show in that case how the analogous results cited above arise.
Then we return to the slightly more involved case of Feshbach resonance
and determine the general phase diagram for this case.
We first provide the results for the case where the system can be
described by an effective single channel model,
i.e., purely in terms of atoms interacting with each other
via a two-body instantaneous interaction.
Then we study the necessary modifications to these results
when this approximation is relaxed.
Finally a probe of the pairing symmetry based on the time-of-flight
experiment is discussed.

\section{II. Lessons from Anisotropic BCS Model}

In the spirit of the work of Anderson and Morel \cite{Anderson61},
we first consider a weak-coupling BCS model of
 a Fermi system with one spin species under the
interaction $-V_{\vec k - \vec k'}$.
(We shall however study the case where this $V$ is anisotropic,
see below).
The Hamiltonian of our system
is $H=H_f + H_V$, where
\begin{eqnarray}
H_f = \sum_{\vec k} (\epsilon_k-\mu )
 a_{\vec k}^{\dagger} a_{\vec k}
\end{eqnarray}
\begin{eqnarray}
H_V = - \sum_{\vec k,\vec k',\vec q} V_{\vec k-\vec k'}
a_{\vec q+\vec k'}^\dagger a_{\vec q-\vec
k'}^\dagger
a_{\vec q-\vec k} a_{\vec q+\vec k} \nonumber \\
\end{eqnarray}
Here $\epsilon_k = \hbar^2 k^2/2M$, and $\mu$ is the chemical potential,
and $a_{\vec k}$ is the annihilation operator for
a Fermion with wave-vector $\vec k$.
 Standard BCS theory proposes the
following trial ground state wavefunction
\begin{eqnarray}
|G\rangle = \prod_k (u_{\vec k} + v_{\vec k} a_{\vec
k}^\dagger a_{-\vec k}^\dagger) |0\rangle
\end{eqnarray}
with normalization condition $|u_{\vec k}|^2 + |v_{\vec k}|^2 =1$.
Minimizing the ground state energy $\langle G|H|G\rangle$ with
respect to $u_{\vec k}$ and $v_{\vec k}$ gives the excitation
spectrum
\begin{eqnarray}
E_{\vec k} = \sqrt{(\epsilon_k-\mu)^2+|\Delta_{\vec k}|^2}
\end{eqnarray}
where the energy gap $\Delta_{\vec k}$ satisfies
\begin{eqnarray} \label{gap1}
\Delta_{\vec k} = \sum_{\vec k'}V_{\vec k-\vec
k'}\frac{\Delta_{\vec k'}}{2E_{\vec k'}}
\end{eqnarray}
In case the interaction is $p$-wave, only the $l=1$ component
survive in $V_{\vec k-\vec k'}$. We write
\begin{eqnarray} \label{vk}
V_{\vec k-\vec k'} = 2\pi  \sum_{m=-1}^{m=1} V_{m}(k,k')
Y_{1m}(\hat k) Y_{1m}^*(\hat k')
\end{eqnarray}
We shall study the situation where
$V_1 = V_{-1} \le V_0$.
We also decompose the energy gap in spherical harmonics of $l=1$
partial waves,
\begin{eqnarray} \label{dk}
\Delta_{\vec k} &=& \Delta f(\hat k) \nonumber \\ &=& \Delta
\sum_{m=-1}^{m=1} f_{m} Y_{1m}(\hat k)
\end{eqnarray}
with normalization $\sum_{m} |f_{m}|^2 =1$. Substituting
Eqs.(\ref{vk})-(\ref{dk}) into Eq.(\ref{gap1}), we have
\begin{eqnarray} \label{gap2}
f(\hat k) = \pi \sum_{\vec k'} \sum_{m} V_{m}(k,k')Y_{1m}(\hat k)
\frac{Y_{1m}^*(\hat k')f(\hat k')}{[(\epsilon_{k'}-\mu)^2+\Delta^2
|f(\hat k')|^2]^\frac{1}{2}}
 \nonumber \\
\end{eqnarray}
Suppose further that $V_{m}(k,k')=V_{m}$ is independent of $k$ and
$k'$, Eq.(\ref{gap2}) becomes
\begin{eqnarray}
f_{m} = \pi V_{m} \sum_{\vec k} \frac{Y_{1m}^*(\hat
k)f(\hat k)}{[(\epsilon_{k}-\mu)^2+\Delta^2 |f(\hat
k)|^2]^\frac{1}{2}}
\end{eqnarray}
We perform the integration in $\vec k$ in the following way,
\begin{eqnarray}
\sum_{\vec k} \rightarrow \rho_0 \int_{-\Lambda}^\Lambda d\epsilon
\int d\Omega
\end{eqnarray}
where $\rho_0$ is the density of state around the Fermi surface,
$\Lambda$ is the energy cutoff, and $d\Omega$ is solid angle. Then
we get totally 3 nonlinear equations,
\begin{eqnarray} \label{gm}
f_{m} = 2\pi \rho_0 V_{m} \left[
 f_{m} \log(\frac{2\Lambda}{\Delta})
 - \int d\Omega Y_{1m}^*(\hat k)f(\hat k)\log|f(\hat k)|
 \right] \nonumber \\
\end{eqnarray}
for $m=-1,0,1$.

 We would like to find the solution
to Eq. (\ref{gm}) with the anisotropy
 characterized by the ratio $V_1/V_0$.
Particular solutions to
Eq.(\ref{gm}) can be easily found at two limits.
When $V_1/V_0=0$,
 $\Delta_{\vec k} \propto Y_{10}(\hat
k)\propto k_z$ because no other pairing can be possible.
 At the
other limit, $V_1/V_0=1$,
 the system is again isotropic the ground
state pairing is $\Delta_{\vec k} \propto Y_{11}(\hat k)$,
or its rotations.
In terms of the language by
$f_m$'s, for $V_1/V_0=0$,
$f_0=1, f_1=f_{-1}=0$ (up to a gauge transformation).
For $V_1/V_0=1$,  a particular solution is
$f_0=1/\sqrt{2}, f_1=f_{-1}=1/2$ such that
\begin{eqnarray}
\Delta_{\vec k} &\propto& \frac{1}{\sqrt{2}}Y_{10}(\hat k)
 + \frac{1}{2}(Y_{11}(\hat k) + Y_{1,-1}(\hat k)) \nonumber \\
 &\propto& k_z - i k_y
\end{eqnarray}
Note that it does not contradict with the pairing $Y_{11}(\hat
k)\propto k_x + i k_y$ since these states are degenerate in an isotropic
limit $V_1=V_{-1}=V_0$.
With
gradually tuning the parameter $0< V_1/V_0 < 1$, we will show
below that $\Delta(\hat k) \propto k_z - i\beta k_y$ up to a
$z$-axis rotation with $\beta < 1$.

Before solving the gap equation
Eq.(\ref{gm}) for the general intermediate values
of the pairing $0< V_1/V_0 < 1$, it is helpful to first
identify the critical $V_1^*/V_0$ such that $f_1$ and $f_{-1}$
start to deviate from zero (i.e., when the pairing starts to deviate from
$Y_{10}$).  Let us take the gauge where $f_0$ is real.
Linearizing Eq.(\ref{gm}) in $f_1$ and $f_{-1}$
around the point $(f_0,f_1,f_{-1})=(1,0,0)$, we have
\begin{eqnarray}
\frac{1}{2\pi\rho_0 V_1^*} \left( \begin{array}{c}
 f_1 \\ f_{-1}^* \end{array} \right) =
\left( \begin{array}{cc}
A_1 & A_2 \\
A_2 & A_1 \end{array} \right)
 \left( \begin{array}{c}
 f_1 \\ f_{-1}^* \end{array} \right)
\end{eqnarray}
where $A_1= \log(2\Lambda/\Delta) -\int d\Omega |Y_{11}|^2
\log|Y_{10}|-1/2$, $A_2=1/2$. Thus when we increase $V_1$, the
first non-trivial solution for $f_1$ and $f_{-1}$ is obtained when
$(f_1, f_{-1}^*)$ gives the  largest eigenvalue of the matrix.  It
can be easily seen that this corresponds to the solution $f_1 =
f_{-1}^*$, with eigenvalue $A_1 + A_2$.  The
critical value $V_1^{*}$ thus satisfies
\begin{equation}
\frac{1}{ 2\pi\rho_0 V_1^*} =  \left[ \log(\frac{2\Lambda}{\Delta})
 - \int d\Omega |Y_{11}|^2 \log|Y_{10}| \right]
 \\ \label{gm1}
\end{equation}
On the other hand, at this point (from Eq. (\ref{gm}) with $m=0$),
\begin{equation}
\frac{1}{ 2\pi\rho_0 V_0} = \left[ \log(\frac{2\Lambda}{\Delta})
 - \int d\Omega |Y_{10}|^2 \log|Y_{10}| \right]
 \label{gm2}
\end{equation}
Eliminating $\log(2\Lambda/\Delta)$ in Eqs.(\ref{gm1})-(\ref{gm2})
gives
\begin{eqnarray}
V_1^*/V_0 = \frac{1}{1+2\pi\rho_0 V_0}
\end{eqnarray}

For $V_1^*/V_0 < V_1/V_0 < 1$, $f_{\pm 1}$ becomes finite.
Provided no additional phase transitions occur,
we expect then the relation $f_1 = f_{-1}^*$ remains valid.
Furthermore,
our system is rotationally invariant around $\hat z$. Under this
rotation, $f_1 \to f_1 e^{i \phi}$ but $f_{-1} \to f_{-1} e^{- i
\phi}$ where $\phi$ is the rotational angle.  By suitable choice
of $\phi$ we can thus make both $f_1$ and $f_{-1} = f_1^*$ real
and positive. We shall confine ourselves to this case without loss
of generality.  Under this choice the order parameter has
the form $k_z - i \beta k_y$.

\begin{figure}[tbh]
\begin{center}
\includegraphics[width=3in]{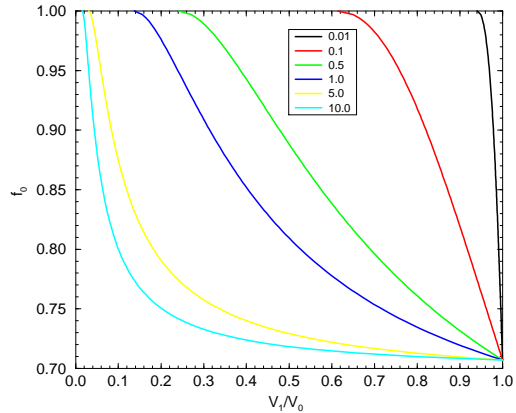}
\end{center}
\vspace{-5pt}
 \caption{$f_0$ as a function of $V_1/V_0$ for $\rho_0 V_0$'s
given in the legend.}
 \label{f0}
 \vspace{-5pt}
\end{figure}

Hence, when $V_1^*/V_0 < V_1/V_0 < 1$, we search for the solution $f_0>0,
f_1=f_{-1}>0$ under the normalization condition $f_0^2 + 2 f_1^2 =
1$. The solution $f_0$ (where $f_1=f_{-1}=\sqrt{(1-f_0^2)/2}$ )
for different $V_1/V_0$ and  $\rho_0 V_0$ is
presented in Fig.\ref{f0}. For given $\rho_0 V_0$,
the pairing is $k_z$ ($f_0=0$) at small $V_1/V_0$ until the latter
exceeds the critical $V_1^*/V_0$. Beyond this critical $V_1^*/V_0$,
$f_0$ decreases gradually with increasing $V_1/V_0$. At
$V_1/V_0=1$, $f_0=1/\sqrt{2}$ for all $\rho_0 V_0$'s meaning that
the pairing becomes $k_z - i k_y$ (corresponding to $\beta = 1$).

\section{III. P-WAVE FESHBACH PAIRING}

We return to the case of p-wave Feshbach resonance.
We begin with the total Hamiltonian $H = H_f + H_b + H_{\alpha}$
where again $H_f$ is for the free fermions,
\begin{eqnarray}  \label{Hf}
H_f &=& \sum_{\vec k} (\epsilon_k  - \mu)
 a_{\vec k}^{\dagger} a_{\vec k} \ .
\end{eqnarray}
$H_b$ is for the molecules in the ``closed" channels, with $b_{\vec q, m}$
the annihilation operator for a Boson with angular momentum $l =
1$ and $z$-axis projection $m$ with momentum $\vec q$,
\begin{eqnarray}
H_b &=& \sum_{\vec q} \left(
    \frac{\epsilon_q}{2} + \delta_m  - 2 \mu \right)
   b_{\vec q, m}^{\dagger} b_{\vec q, m}
\label{Hb}
\end{eqnarray}
Here $\delta_m$'s are the (bare) detuning of the $l=1$, $m$ resonance.
$H_\alpha$ represents the Feshbach coupling, which is a
generalization of the ones already commonly employed
\cite{ohashi,Fesh} for $s$-wave Feshbach
resonances to the case of several $l =1$, $m = 0, \pm 1$ closed
channels \cite{ohashi05,ho05}:
\begin{eqnarray} \label{Hal}
H_{\alpha} =
   \sum_{m,\vec q, \vec k}
   \varphi^*_m(\vec k)
      b_{\vec q, m}^\dagger a_{- \vec k + \vec q/2} a_{\vec k + \vec q/2}
   + {\rm h.c.}
\end{eqnarray}
where $\varphi_m(\vec k)=-\frac{i \sqrt{ 4 \pi}}{L^{3/2}}  k Y_{1
m}(\hat k) \alpha_m$. The factor $Y_{1 m}(\hat
k)$ reflects the symmetry of the $l, m$ bound state and the linear
factor in $k$ arises from the small momentum approximation for the
coupling. $\alpha_m$ is the corresponding coupling constant,
it has dimension $[E][L]^{5/2}$.  More precisely $\alpha_m$ can be
an analytic function of $k$.  For low energy (density) phenomena,
we shall however only need its value $\alpha_m \equiv \alpha_m(k=0)$.

We shall apply mean-field theory to the Hamiltonian $H$.
In this approximation, we discard all terms that involve
finite momentum Bosons $b_{\vec q, m}$ with $\vec q \ne 0$,
and regard $b_{\vec q = 0, m}$ as c-numbers.
It is then convenient to introduce
\begin{eqnarray}
D_m &=& - \frac{i\sqrt{4\pi}}{L^\frac{3}{2}} \alpha_m b_{0,m}  \label{defDm}
\\
\Delta_{\vec k} &=& \sum_m D_m k Y_{1 m}(\hat k) \label{defDelta}
\end{eqnarray}
Hence the mean-field Hamiltonian for the Fermions becomes
\begin{eqnarray} \label{hmf}
H^{\rm mf} = \sum_{\vec k} (\epsilon_k -\mu)a_{\vec k}^\dagger
a_{\vec k} + (\Delta_{\vec k}^* a_{-\vec k}a_{\vec k} + {\rm
h.c.})
\end{eqnarray}
This can be solved via the familiar
Bogoliubov transformation, which gives
\begin{eqnarray}
\langle a_{-\vec k}a_{\vec k} \rangle =
 - \frac{\Delta_{\vec k}}{2[(\epsilon_k-\mu)^2+|\Delta_{\vec
k}|^2]^\frac{1}{2}}
\end{eqnarray}
Minimization with respect to the c-numbers $b_{\vec q = 0, m}$ give
\begin{eqnarray}
b_{0,m} (\delta_m - 2\mu) = - \sum_{\vec k}
 \varphi^*_m(\vec k) \langle a_{-\vec k}a_{\vec k} \rangle
\end{eqnarray}
which can be written as,
\begin{eqnarray} \label{Dm}
(\delta_m -2 \mu ) D_m
  = \frac { 2 \pi} {L^3}  \sum_{\vec k}
   \frac {|\alpha_m|^2
    k   Y_{1 m}^*(\hat k) \Delta_{\vec k}} { \left[
(\epsilon_k-\mu)^2+|\Delta_{\vec k}|^2
\right]^\frac{1}{2}  }
\label{gap-un}
\end{eqnarray}

The R.H.S. of this equation is formally divergent
(This divergence is only formal:  the sum may actually
converge if the $k$ dependences of $\alpha_m(k)$
are included)
Nevertheless,
it can be "renormalized" by re-expressing this equation
in terms of parameters that enter the
two-body scattering amplitude
(c.f. \cite{ho05} and also the theoretical references
in the s-wave case \cite{cross0,falco04,perali,java,numeric}.)
Let us then consider the scattering of two particles in vacuum.
We denote $f_{\vec k}(\vec k')$ as the
scattering amplitude for
two incident particles with relative momentum $\vec k$ and
out-going momentum $\vec k'$.
We can define
 $f_m(k)$ (do not confuse
this $f_m(k)$ with those in Sec II)
via
\begin{equation}
f_{\vec k} (\vec k') = \sum_m  f_m(k)
(4 \pi) Y_{1m}(\hat k') Y_{1m}^{*} (\hat k)
\label{fmdef}
\end{equation}
This equation is the generalization of the corresponding
one in scattering theory to the case of anisotropic
interaction (but still with rotational invariance around $\hat z$).
  It reduces to those in standard
quantum mechanics textbooks \cite{Merzbacher}
if $f_m(k)$ is independent of $m$.
$f_{\vec k}(\vec k')$ is related
to the $T$-matrix by
\begin{eqnarray} \label{flm}
f_{\vec k}(\vec k') = -\frac{M L^3}{4\pi \hbar^2}
T_{\vec k',\vec k}(E=2\epsilon_k + i0^+)
\end{eqnarray}
The $T$-matrix can be evaluated in the standard way:
\begin{eqnarray} \label{tmatrix}
T_{\vec k',\vec k}(E) &=& \sum_m \big[ \varphi_m(\vec k'){\cal
G}^{(m)}_{b_0}(E)\varphi^*_m(\vec k) \nonumber \\
 &&+ \varphi_m(\vec k'){\cal G}^{(m)}_{b_0}(E)\Pi^{(mm)}(E) {\cal
G}^{(m)}_{b_0}(E)\varphi_m^*(\vec k) + \cdots \big] \nonumber \\
 &=& \sum_m \frac{\varphi_m^*(\vec k')\varphi_m(\vec k)}
              {E-\delta_m-\Pi^{(m)}(E)}
\end{eqnarray}
where
\begin{eqnarray} \label{pi}
\Pi^{(mm')}(E) &=& \sum_{\vec p,\omega} \varphi_m^*(\vec p) {\cal
G}_{f}(\vec p,\omega) {\cal G}_{f}(-\vec p, E - \omega)
\varphi_{m'}(\vec p) \nonumber \\
 &=& \sum_{\vec p} \frac{\varphi_m^*(\vec p)\varphi_{m'}(\vec p)}
    {E-2\epsilon_p}   \ .
\end{eqnarray}
${\cal G}_{f}(\vec p, \omega)$
is the Greens function for free Fermions
at wavevector $\vec p$ ($p$ has dimension of $[L]^{-1}$),
 frequency $\omega$,
and ${\cal G}^{(m)}_{b_0}(E)$ is the Greens
functions of free  $m$-Boson
  at zero momentum $\vec q = 0$ and energy $E$.
Here we have already made use of a simplification due to the fact
that $\Pi^{(m m')}$ vanishes for $m \ne m'$ (see Eq. (\ref{pi})),
hence there are no cross-terms between different $m$'s in Eq.
(\ref{tmatrix}).

Substituting Eqs.(\ref{tmatrix})-(\ref{pi}) into Eq.(\ref{flm}),
and using the definition (\ref{fmdef}), we get
\begin{eqnarray} \label{high1}
-\frac{M}{4\pi \hbar^2}\frac{k^2}{f_m(k)} =
\frac{2\epsilon_k-\delta_m}{|\alpha_m|^2} -
\frac{1}{L^3}\sum_{\vec p} \frac{p^2}{2\epsilon_k-2\epsilon_p +
i0^+} \nonumber \\
\end{eqnarray}
The sum over $\vec p$ in the R.H.S. formally diverges.
However, we are interested only on its dependence
on $k$ at small $k$'s.  We thus expand the
R.H.S. in powers of $k$:
\begin{eqnarray} \label{high2}
&& \frac{2\epsilon_k-\delta_m}{|\alpha_m|^2} -
\frac{1}{L^3}\sum_{\vec p} \frac{p^2}{2\epsilon_k-2\epsilon_p +
i0^+} \nonumber \\
 &=& -( \frac{\delta_m}{|\alpha_m|^2}
   -  \frac{M}{\hbar^2} \frac{1}{L^3} \sum_{\vec p} 1  ) \nonumber \\
 &&  + (  \frac{\hbar^2}{ M |\alpha_m|^2}
   +  \frac{M}{\hbar^2} \frac{1}{L^3} \sum_{\vec p} \frac{1}{p^2} )k^2 -
   i\frac{M}{4\pi \hbar^2}k^3
\end{eqnarray}

\vspace{15pt}
\begin{figure}
\begin{center}
\includegraphics[width=3in]{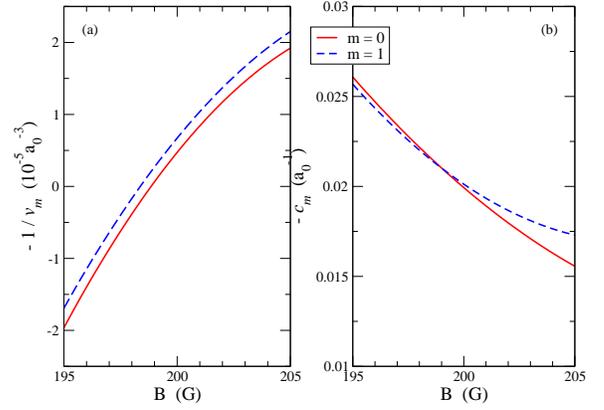}
\end{center}
 \caption{Plot of the fit to the experimental results
of Ref.\cite{Ticknor04} for $-1/v_m$ and $-c_m$ for $^{40}K$ atoms in the
$ |f, m_f\rangle = | 9/2, -7/2\rangle$ hyperfine state.
$v_1 = v_{-1}$ and $c_1 = c_{-1}$.
 $-1/v_m$'s vanish at the resonant fields $B_m^*$, with
$B_0^* > B_{\pm 1}^*$.}
 \label{mag}
\end{figure}


In the low-energy limit, Eq.(\ref{high1})-({\ref{high2}) should
reduce to the general result
 for the low-energy scattering between a pair
of particles.  Using the same notation adopted in
Ref.\cite{Ticknor04}, the scattering amplitude is parametrized as
\begin{eqnarray} \label{high3}
f_m(k) = \frac{k^2}{ -v_m^{-1}+ c_m k^2 - i k^3} \label{fm}
\end{eqnarray}
where the dimensions of $v_m$ and $c_m$ are $\rm [L]^3$ and $\rm
[L]^{-1}$, respectively.
The magnetic field dependent parameters
$v_m$, $c_m$ are in principle available experimentally.
An example is as shown in  Fig.\ref{mag}.
 Our renormalization scheme is to absorb the bare
parameters $\delta_m$ and $\alpha_m$ into the physical
(renormalized) parameters $v_m$, $c_m$ by identifying the
equivalence between Eqs.(\ref{high1})-(\ref{high2}) and
Eq.(\ref{high3}). Hence
\begin{eqnarray}
- \frac{1}{v_m} &=& \frac{ 4 \pi \hbar^2}{M} (
  \frac{\delta_m}{|\alpha_m|^2}
   - \frac{M}{\hbar^2} \frac{1}{L^3} \sum_{\vec p} 1 )
\label{vm} \\
c_m &=& - \frac{ 4 \pi \hbar^2}{M} (
  \frac{\hbar^2}{ M |\alpha_m|^2}
   +  \frac{M}{\hbar^2} \frac{1}{L^3} \sum_{\vec p} \frac{1}{p^2} )
\label{cm}
\end{eqnarray}
 In Eq. (\ref{vm}) and (\ref{cm}), if the $p$ dependence
of $\alpha_m$'s had been included, the sums involving $\vec p$
should have the extra factors $ |\alpha_m(p)|^2/ |\alpha_m(0)|^2$.
When treating the many-body problem, as we shall see below, the
only "divergent" sums are again $ \sum_{\vec p} |\alpha_m(p)|^2/
|\alpha_m(0)|^2$ and $ \sum_{\vec p} |\alpha_m(p)|^2/
|\alpha_m(0)|^2 p^2$. We shall apply Eq. (\ref{vm}) and (\ref{cm})
to eliminate these sums in favor of $v_m$ and $c_m$. After this
procedure, low energy phenomena (involving only small $p$'s) can
be parameterized entirely by $v_m$, $c_m$ and $\alpha_m(0)$. To
simplify our presentation, we shall not include
 explicitly the factors
$ |\alpha_m(p)|^2/ |\alpha_m(0)|^2$
in our calculations below.

Returning to Eq.(\ref{Dm}), we rewrite its R.H.S as
\begin{eqnarray}
&& \frac{2\pi}{L^3}\sum_{\vec k}
 \frac{|\alpha_m|^2 k Y_{1 m}^*(\hat k) \Delta_{\vec k}}
 {\left[ (\epsilon_k-\mu)^2+|\Delta_{\vec k}|^2
 \right]^\frac{1}{2}} \nonumber \\
 &=& \frac{2\pi|\alpha_m|^2}{L^3}\sum_{\vec k,m'}
\frac{D_{m'} Y_{1 m}^*(\hat k) Y_{1 m'}(\hat k) k^2}
 {\left[ (\epsilon_k-\mu)^2+|\Delta_{\vec k}|^2
 \right]^\frac{1}{2}} \nonumber \\
&=& \frac{2\pi|\alpha_m|^2}{L^3}\sum_{\vec k,m'}D_{m'} Y_{1
m}^*(\hat k) Y_{1 m'}(\hat k)
  \nonumber \\
 && \times \left\{ h(\vec k) +
\frac{k^2}{\epsilon_k}(1+\frac{\mu}{\epsilon_k}-\frac{|\Delta_{\vec
k}|^2}{2\epsilon_k^2}) \right\}
\label{rngap}
\end{eqnarray}
where
\begin{eqnarray}
h(\vec k) \equiv \frac{k^2}{\left[ (\epsilon_k-\mu)^2 +
|\Delta_{\vec k}|^2 \right]^{1/2} }
 - \frac{k^2}{\epsilon_k}(1+\frac{\mu}{\epsilon_k}-\frac{|\Delta_{\vec
k}|^2}{2\epsilon_k^2}) \nonumber \\
\label{hk}
\end{eqnarray}
The sum involving $h(\vec k)$ in Eq. (\ref{rngap}) is convergent.
For the rest of the terms, their divergences are the same as in
either Eqs. (\ref{vm}) and (\ref{cm}). These sums can thus be expressed
in terms of the physical parameters $v_m$ and $c_m$.
Using Eq.(\ref{rngap}), the "gap equation" Eq.(\ref{Dm})
 can therefore be rewritten as
\begin{widetext}
\begin{eqnarray}
 \frac{M}{4\pi}D_0(-\frac{1}{v_0}+\frac{2Mc_0\mu}{\hbar^2})
 - \frac{3M}{10\pi}(\frac{M^2 c_0}{4\pi\hbar^4}+\frac{1}{|\alpha_0|^2})
 \left[D_0(2|D_1|^2+3|D_0|^2+2|D_{-1}|^2)-2D_{0}^*D_1 D_{-1}
   \right] \nonumber \\
 = \frac{2\pi}{L^3}\sum_{\vec k, m}D_m Y_{10}^*(\hat k) Y_{1
m}(\hat
 k) h(\vec k)
 \label{D0}
\end{eqnarray}
\begin{eqnarray}
 \frac{M}{4\pi}D_1(-\frac{1}{v_1}+\frac{2Mc_1\mu}{\hbar^2})
 - \frac{3M}{10\pi}(\frac{M^2 c_1}{4\pi\hbar^4}+\frac{1}{|\alpha_1|^2})
 \left[ D_1
(3|D_1|^2+2|D_0|^2+6|D_{-1}|^2)-D_{-1}^*D_0^2 \right] \nonumber \\
 = \frac{2\pi}{L^3}\sum_{\vec k, m}D_m Y_{11}^*(\hat k) Y_{1 m}(\hat
 k) h(\vec k)
 \label{D1}
\end{eqnarray}
\end{widetext}
and a corresponding equation with $m= 1 \leftrightarrow m=-1$.
Eq.(\ref{D0})-(\ref{D1}) are to be solved under the constraint from the
number equation
\begin{eqnarray}
n = \frac{1}{L^3} ( \sum_{\vec k} \langle a_{\vec
k}^{\dagger} a_{\vec k} \rangle
   + 2 \sum_m |b_{0,m}|^2 )
\label{n1}
\end{eqnarray}
The first term is from open-channel atoms whereas
the second term is from the closed-channel molecules.
From BCS theory,
\begin{displaymath}
\langle a_{\vec k}^{\dagger} a_{\vec k} \rangle
= \frac{1}{2} \left( 1-\frac{\epsilon_k - \mu}{[
(\epsilon_k-\mu)^2+|\Delta_{\vec k}|^2]^\frac{1}{2}} \right)
\end{displaymath}
 The first term in Eq. (\ref{n1}) is thus again formally divergent:
 $ \langle a_{\vec k}^{\dagger} a_{\vec k} \rangle$ $\sim$
$ \frac{ |\Delta_{\vec k} |^2} { 4 \epsilon_k^2 }$ at large $k$.
 However, this sum
it can be treated in analogous manner as Eq. (\ref{rngap}) by
employing again Eq. (\ref{cm}).
 The result is
\begin{eqnarray}
n &=& \frac{1}{2L^3} \sum_{\vec k} ( 1-\frac{\epsilon_k - \mu}{[
(\epsilon_k-\mu)^2+|\Delta_{\vec k}|^2]^\frac{1}{2}}
           - \frac{ |\Delta_{\vec k} |^2} { 2 \epsilon_k^2 } )
           \nonumber \\
  && - \frac{1}{4\pi}\sum_m(\frac{M^2
c_m}{4\pi\hbar^4}+\frac{1}{|\alpha_m|^2})|D_m|^2
   + \frac{1}{2\pi}\sum_m
  \frac{|D_m|^2}{|\alpha_m|^2}  \nonumber \\
\label{n}
\end{eqnarray}

Eqs.(\ref{D0}), (\ref{D1}), and (\ref{n}) are our principal
equations, with parameters characterizing the Feshbach resonances
entirely in terms of $v_m$, $c_m$ and $1/|\alpha_m|^2$.
 These equations determine
the order parameters $D_m$ and chemical potential $\mu$ for given
density $n$ and ``interaction parameters" $v_m$, $c_m$ and
$1/|\alpha_m|^2$.

\section{IV. BEC-BCS crossover and quantum phase transitions}

Generally with Feshbach resonance for the sub-channel $m$, $1/v_m$ is field
dependent, vanishing at the resonant field $B_m^{*}$.  In contrast, $c_m$ has
a definite sign ({\it c.f.} Fig. \ref{mag}).
 For the ease of discussions, we shall assume that
$c_m < 0$ and field independent, $-1/v_m$  is an increasing
function of field
  $- 1/v_m > (<) 0$
for $B > (<) B_m^{*}$, as in the case of $^{40}$K.
 (This corresponds to the case where
$\delta_m$ is an increasing function of field and $\alpha_m$
weakly field dependent, {\it c.f.} Eqs.(\ref{vm}) and (\ref{cm})).
For $B < B_m^{*}$, a bound state appears.  The energy of this
bound state is given by $- \epsilon_{b,m} = - \hbar^2
{\kappa_m}^2/M$ with $k = i \kappa_m$ being a pole for $f_m (k)$,
that is, $ - 1/v_m - c_m \kappa_m^2 - \kappa_m^3 = 0$.
For given $m$ and small detuning below that resonance,
$\kappa_m$ is small and is given by $\kappa_m^2 = 1/ [
(-c_m)(v_m) ]$.  Since $1/v_m$ should be roughly linear in $B$
near the resonance, $\epsilon_{b,m}$ increases linearly with $
(B_m^{*} - B)$ (in contrast to $s$-wave, where it is quadratic;
see also \cite{ho05})
These results for small detuning apply provided $\kappa_m << (-c_m)$,
or equivalently $1/v_m \ll (-c_m)^3$.
We shall always confine ourselves to this regime for
negative detunings (even when we write $ B \ll B_m^*$).
For larger negative detunings we need to take into account
higher order terms in $k^2$ in denominator of $f_m(k)$.

Moreover, as explained in Ref.\cite{Ticknor04}, due to the dipole
interaction, $B_0^{*} > B_{1}^{*} = B_{-1}^{*}$,
as can be seen again in Fig \ref{mag}.
Thus in the field
range of interest, $ - 1/v_1 = -1/v_{-1} > -1/v_0$. We can say
that, at a given field, the effective interaction between the
Fermions is less attractive for relative angular momentum
projections $m = \pm 1$ than $m = 0$.

Since the interaction is less attractive for angular momentum
projections $m = \pm 1$, for sufficiently large difference between
$-1/v_{0}$ and $-1/v_{\pm 1}$ we expect (and verify below) that
the pairing is entirely in the $m = 0$ partial wave.  We thus
first begin our analysis by assuming that only $D_0$ is
non-vanishing.

\vspace{25pt}
\begin{figure}[tbh]
\begin{center}
\includegraphics[width=3in]{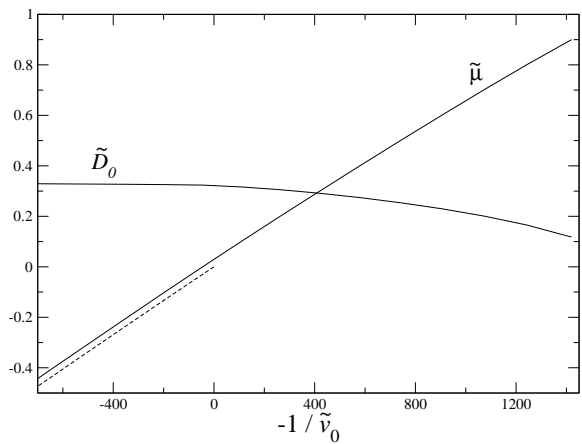}
\end{center}
 \caption{The dimensionless parameters $\tilde D_0$ and $\tilde
 \mu$ as functions of $-1/\tilde v_0$.
$\tilde c_0= \tilde c_1=-100$ and $\tilde D_{\pm 1} =0$ in this
case. The dashed line represents $ - \epsilon_b / 2 \epsilon_F$.
  }
 \label{gap_c100}
\end{figure}

For simplicity, we shall first drop the terms with explicit $1 /
|\alpha_m|^2$ factors. If this is valid, then our equations
(\ref{D0}), (\ref{D1}), (\ref{n}) are same as a corresponding
model consisting only of Fermions interacting among each other
with an interaction such that the corresponding
 two-body scattering amplitude parameters are given
by the same $c_m$ and $v_m$.
In analogy with the s-wave literature, we shall call this the
single-channel approximation
(note there are still three  ($m = -1, 0, 1$) Feshbach resonances).
 The effect from finite $1/|\alpha_m|^2$ will be discussed at the end
of this Section.

With these simplifications, Eq. (\ref{D0}) and (\ref{n}) become
\begin{equation}
 \frac{M}{4\pi}(-\frac{1}{v_0}+\frac{2Mc_0\mu}{\hbar^2})
 + \frac{9M}{40\pi^2\hbar^4}(-c_0)|D_0|^2
 = \frac{2\pi}{L^3}\sum_{\vec k, m} |Y_{10}(\hat k)|^2 h(\vec k)
 \label{D00}
\end{equation}
and
\begin{eqnarray}
n &=& \frac{1}{2L^3} \sum_{\vec k} ( 1-\frac{\epsilon_k - \mu}{[
(\epsilon_k-\mu)^2+|\Delta_{\vec k}|^2]^\frac{1}{2}}
           - \frac{ |\Delta_{\vec k} |^2} { 2 \epsilon_k^2 } )
           \nonumber \\
  && + \frac{M^2}{(4\pi)^2\hbar^4} (-c_0) |D_0|^2
\label{n00}
\end{eqnarray}
respectively.
 Eqs.(\ref{D0}) and (\ref{n}) can be solved
simultaneously similar to the $s$-wave case.
By gauge invariance, we shall choose $D_0$ to be real
without loss of generality.
It is convenient to express the results in
dimensionless form. We thus define $\tilde \mu\equiv \mu/\epsilon_{\rm
F}$, $\tilde D_m \equiv D_m/ \hbar v_{\rm F}$, $\tilde c_m \equiv
n^{-1/3}c_m$, and $\tilde v_m \equiv n v_m$ where $\epsilon_{\rm
F} \equiv \hbar^2 k^2_{\rm F}/2M$, $v_{\rm F} \equiv \hbar k_{\rm F}/M$,
and $k_{\rm F}^3 \equiv 6\pi^2 n$.
The results are as shown in
Fig. \ref{gap_c100} (for the case $\tilde c_0 = \tilde c_1 = -
100$, see below for the reason of this choice).

In the BCS regime, corresponding to ``large" external magnetic
field $B\gg B^*_0$ thus $-{\tilde v_0}^{-1} \to \infty$, we have
$\tilde \mu \simeq 1$, and $\tilde D_0\simeq 0$.
We can ignore the term $\propto |D_0|^2$
 on the left hand side of Eq.(\ref{D00}).
 This equation then
reduces to the corresponding one in Sec II with the effective
pairing interaction $V_0$ in the $m=0$ channel there
proportional to $ (-1/v_0 + 2M c_0 \mu/\hbar^2 )^{-1}$ $=$
$(-1/v_0 + c_0 k_F^2)^{-1}$.
This combination follows from the factor
 $(\delta_0 - 2 \mu)$ on the left hand side of Eq. (\ref{gap-un}).

In the BEC regime,  corresponding to $B\ll B^*_0$ or $-{\tilde
v}_0^{-1}\rightarrow -\infty$, $\mu \simeq -\epsilon_{b,0}/2$, and
$\tilde D_0$ approaches to a constant. In Eq. (\ref{D00}) we can
ignore all terms explicit in $\Delta_{\vec k}$ or $D_0$.  Writing
$\kappa = (2 M |\mu|)^{1/2}$ and performing the sum over $\vec k$,
we find that $\kappa$ obeys $-1/v_0 -c_0 \kappa^2 = \kappa^3$.
Comparing this with the equation for the two body bound state, we
find $|\mu| = \epsilon_{b 0}/2$ as claimed. In Eq. (\ref{n00}), we
can expand the square-root $[(\epsilon_k - \mu) + |\Delta_{\vec
k}|^2]^{-1/2}$ $\approx$ $ (\epsilon_k + |\mu|) - |\Delta_{\vec
k}|^2/ 2 (\epsilon_k + |\mu|)$. We then get
\begin{equation}
n = \frac{M^2}{(4\pi)^2 \hbar^4} |D_0|^2 \left[ (-c_0) - \frac{3}{2}\kappa
\right]
\label{n0bec}
\end{equation}
where $\kappa$ was defined above.  For small detuning
$\kappa \ll (-c_0)$, {\it i.e.} $1/\tilde v_0 \ll |\tilde c_0|^3$,
a relation well-satisfied in the range in Fig \ref{gap_c100}
and others below, we then obtain
 $\tilde D_0 \simeq (32 \pi/3)^{1/3}
(- \tilde c_0)^{-1/2}$.

The ``cross-over" behavior in Fig.\ref{gap_c100} is analogous to
the $s$-wave case, where the corresponding $x$-axis is $ x = - 1/
(n^{1/3} a_s)$ where $a$ is the $s$-wave scattering length.
 Note here
$D_0$ has the dimension of $\rm [E][L]^{-1}$ and
behaves differently from the $s$-wave $\Delta$ in the BEC limit:
it saturates rather than continue to increase.
Rigorously speaking, we actually expect a (quantum) phase transition
at the point where $\mu$ crosses zero \cite{randeria90,klinkhamer}:
the system is gapless for $\mu > 0$ but gapful if $\mu < 0$.
However, we found no evident changes of slope
in $\mu$ and $\Delta$ when this transition is crossed,
and thus expect this phase transition is very weak thermodynamically.
Similar results were found in Ref \cite{ho05} for the axial
($D_1 \ne 0$, $D_0 = D_{-1} = 0$ state).

We have also performed calculations for other values of $\tilde
c_0$. The size of the crossover region is roughly proportional to
the value of $\tilde c_0$ (which is in turn inversely proportional
to the density).  For example, for $\tilde c_0 = -200$ (not shown), the
corresponding results can be captured well by replacing the
$x$-axis by $ - 1/ 2 \tilde v_0$ and dividing $\tilde D_0$ by $1 /
\sqrt{2}$ in Fig.\ref{gap_c100}.

The above behavior applies only to sufficiently large $ -\tilde
v_{\pm 1}^{-1} - ( -\tilde v_0^{-1}) > 0$.  When this difference
is sufficiently small, $\tilde D_{\pm 1}$ will become finite.
Similar to the treatment in Section II for the
anisotropic weak-coupling BCS model,
we first identify the critical $\tilde
v_{\pm 1}^{* -1}$ such that $D_1$ and/or $D_{-1}$ first start to deviate from
zero.
For simplicity, in the results below we shall take $c_0 = c_{\pm 1}$.
This is satisfied to a good approximation at least for
$^{40}$K near the resonances (see Fig \ref{mag}).
We linearize
Eqs.(\ref{D0})-(\ref{D1}) (and the latter with
$m=1 \leftrightarrow m=-1$) in $D_1$ and $D_{-1}$.
We obtain
\begin{eqnarray}
-\frac{M}{4\pi v_{1}^*} \left( \begin{array}{c}
 D_1 \\ D_{-1}^* \end{array} \right) =
\left( \begin{array}{cc}
A_1 & A_2 \\
A_2 & A_1 \end{array} \right)
 \left( \begin{array}{c}
 D_1 \\ D_{-1}^* \end{array} \right)
\end{eqnarray}
where
\begin{eqnarray}
A_1 &=& \frac{2\pi}{L^3}\sum_{\vec k} |Y_{11}|^2 h(\vec k)
 -\frac{2\pi}{L^3}\sum_{\vec k} D_0^2 |Y_{10}|^2 |Y_{11}|^2 g(\vec
 k) \nonumber \\
&& - \frac{M^2 c_1}{2\pi \hbar^2}(\mu-\frac{3M}{10\pi \hbar^2}D_0^2) \\
 A_2 &=& \frac{2\pi}{L^3}\sum_{\vec k} D_0^2 |Y_{10}|^2 |Y_{11}|^2
 g(\vec k) - \frac{3M^3 c_1}{40\pi^2 \hbar^4}D_0^2 \\
 g(\vec k) &\equiv& \frac{k^4}{2}(
 \frac{1}{\left[(\epsilon_k -\mu)^2+ D_0^2 |Y_{10}|^2 k_z^2 \right]^{3/2}}
 - \frac{1}{\epsilon_k^3} )
\end{eqnarray}
Here $h(\vec k)$ is as defined in Eq. (\ref{hk}) but with only
$D_0 \ne 0$. We can verify that $A_2 > 0$.  Therefore the largest
eigenvalue for the A-matrix, hence the smallest value for
$v_{1}^*$, belongs to the class $D_1 = D_{-1}^*$. The
corresponding critical value is then given by $ -\frac{M}{4\pi
v_{1}^*} = A_1 + A_2$. On the other hand, from Eq.(\ref{D0}), we
have
\begin{eqnarray}
-\frac{M}{4\pi v_0}
 = \frac{2\pi}{L^3}\sum_{\vec k} |Y_{10}|^2 h(\vec k)
  - \frac{M^2 c_0}{2\pi\hbar^2}(\mu-\frac{9M}{20\pi\hbar^2}D_0^2)
  \nonumber \\
\end{eqnarray} Combining these two equations
and re-writing it in dimensionless form, we have
\begin{eqnarray}
&&-\frac{1}{\tilde v_1^*}+\frac{1}{\tilde v_0} =
\frac{3(6\pi^2)^\frac{2}{3}}{5\pi}{\tilde D_0}^2(-\tilde c_0)
\nonumber \\
&&+ 9\pi \int_0^\infty dx \int_{-1}^1 dy
 \frac{x^4(1-3 y^2)}{[(x^2-\tilde \mu)^2+\frac{3}{\pi}\tilde D_0^2 x^2
y^2]^{1/2}} \nonumber \\
 \label{v*}
\end{eqnarray}

In the BCS limit, $\tilde D_0\rightarrow 0$, the first term in the
R.H.S. of Eq.(\ref{v*}) is negligible whereas the second,
 which we shall call $K_1$, is finite.
The dominant contribution to $K_1$ occurs near $x \simeq \sqrt{\tilde
\mu}$, where the integrand diverges if $\tilde D_0$
were zero exactly. To evaluate $K_1$ in the limit $\tilde D_0 \to
0$, we add and subtract the integral
\begin{equation}
K_2 \equiv 9\pi \int_{-\xi_0}^{\xi_0} d\xi \int_{-1}^1 dy
 \frac{(1-3y^2)}{2[\xi^2+\frac{3}{\pi}\tilde
D_0^2 y^2 ]^\frac{1}{2}}
\label{K2}
\end{equation}
where $\xi \equiv x^2 - \tilde \mu$, and $\xi_0 \gg \tilde D_0$
 is an arbitrary cut-off.  The integrand for
$K_1 - K_2$ now has no divergence near $ x^2 - \tilde \mu \approx 0$,
and their difference vanishes in the $\tilde D_0 \to 0$ limit due
to the $y$ integral.
$K_2$ can be evaluated by first integrating with respect to $\xi$,
we get
\begin{eqnarray}
&& 9\pi  \int_{-1}^1 dy \int_{-\xi_0}^{\xi_0} d\xi
 \frac{(1-3y^2)}{2[\xi^2+\frac{3}{\pi}\tilde
D_0^2 y^2 ]^\frac{1}{2}} \nonumber \\
&=& -18\pi  \int_{-1}^1 dy\ (1-3y^2)\log y
\nonumber \\
&=& 12\pi
\end{eqnarray}
In going from the first to the second line, we have
used the fact that
$ \int_{-1}^1 dy\ (1-3y^2) = 0$, hence the terms involving
$\xi_0$ and $\tilde D_0$ drop out in the  $\xi_0 \gg \tilde D_0$ limit.
Therefore in the BCS limit we obtain $-\tilde
v_1^{* -1} + \tilde v_0^{-1} \rightarrow 12\pi = 37.7$.

In the BEC limit, recalling that $\tilde D_0$ approaches
a constant  $\simeq (32 \pi/3)^{1/3} (- \tilde c_0)^{-1/2}$
whereas $\tilde \mu$ is large and negative,
 we find that the second term in the RHS of Eq. (\ref{v*}) is
negligible compared with the first (provided $\kappa \ll (-c_0)$).
Using the value for $\tilde D_0$, Eq. (\ref{v*}) yields $-\tilde
v_1^{* -1} + \tilde v_0^{-1}\rightarrow 48\pi/5 = 30.2$. In the
intermediate regime, the numerical results for $-\tilde v_1^{* -1}
+ \tilde v_0^{-1}$ as a function of $-\tilde v_0^{-1}$ is shown as
the thick black line ($\tilde D_0 = 0$) in Fig.\ref{beta_c}.



We now solve for the order parameters $D_{m}$ when
$-\tilde v_1^{-1} + \tilde v_0^{-1}$ is less than the critical
value.  As already mentioned, we have already made use of
 gauge invariance to  choose
$D_0$ to be real. Under this choice, the solutions we found belong to the
class $D_{1} = D_{-1}^{*}$. Writing $D_{1} = |D_{1}| e^{i \chi}$,
$\Delta_{\vec k}$ then has the angular dependence $\propto$ $ D_0
\hat k_z + i \sqrt{2} |D_1| \hat k \cdot \hat a$ $\propto (\hat
k_z - i \beta \hat k \cdot \hat a)$ where $\hat a = (\cos \chi)
\hat y + (\sin \chi) \hat x$ is a unit vector perpendicular to
$\hat z$ and $\beta = \sqrt{2} |D_{1} / D_{0}|$. A particular
solution is given by the case where $D_1$ and $D_{-1}$ are both
real where $\hat a = \hat y$. Other solutions are simply
related to this one by a rotation about $\hat z$.  Without loss of
generality, we shall therefore assume that $D_{m}$'s are all
real below.

The contour plot of $\beta$ (by solving Eq. (\ref{D0})-(\ref{n})
without the $1/|\alpha_m|^2$ terms) is shown in Fig.\ref{beta_c}.
(see Fig. 2 of \cite{Cheng05} for a plot of $D_{\pm 1}$) $\beta$
monotonically increases downwards or towards the right in this
diagram. In the $\beta = 0$ ($D_{\pm 1} = 0$) phase, the state is
rotationally invariant about $\hat z$, whereas this symmetry is
broken in the $\beta \ne 0$ phase. There is a (quantum) phase
transition between these two phases when one crosses the critical
line $-\tilde v_1^{* -1} + \tilde v_0^{-1}$.

\vspace{45pt}
\begin{figure}
\begin{center}
\includegraphics[width=3in]{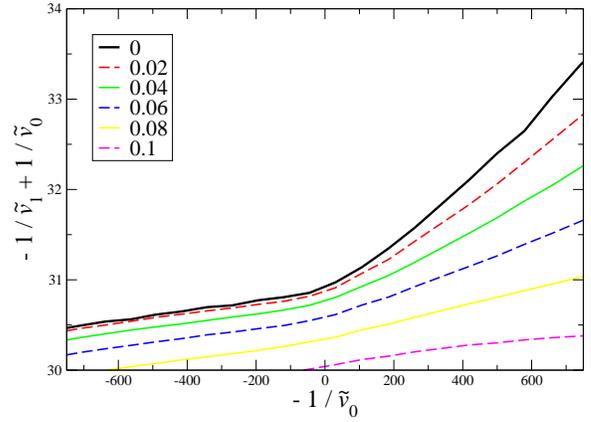}
\end{center}
 \caption{Contour plot of $\beta$ as a function of
$-1/\tilde v_1+ 1/\tilde v_0$ and $-1/\tilde v_0$ for
 $\tilde c_0= \tilde c_{\pm 1}= -100$.
The line for $\beta =0$ corresponds to the critical value
$-1/\tilde v_1^*+ 1/\tilde v_0$.
} \label{beta_c}
 \vspace{15pt}
\end{figure}


For $\tilde
c_0=-200$, the results are qualitatively the same
provided we double the
value of $-1/\tilde v_0$. The corresponding plot is shown in
Fig.\ref{beta_c200}.

In the intermediate splitting regime, we thus predict the state is
$k_z - i\beta k_y$ on the BCS side, whereas it is
$ k_z$ on the BEC side. For large positive detuning, the splitting
should be less relevant and the pairing state should resemble more
that of the isotropic system. On the BEC side, the system should
be closer to a Bose condensate of lowest energy molecules ($\hat
k_z$).

\vspace{75pt}
\begin{figure}
\begin{center}
\includegraphics[width=3in]{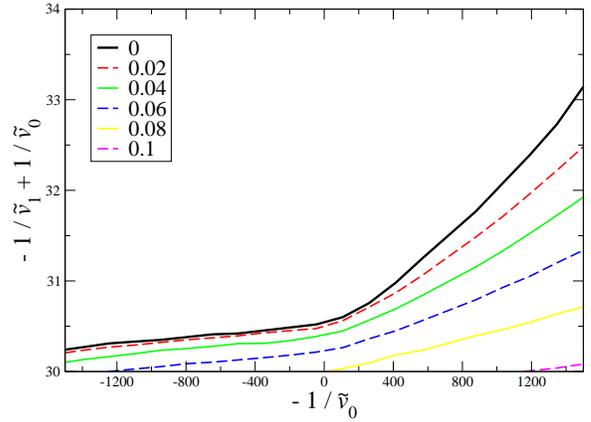}
\end{center}
 \caption{Contour plot of $\beta$ as a function of
$-1/\tilde v_1+ 1/\tilde v_0$ and $-1/\tilde v_0$ for
 $\tilde c_0= \tilde c_{\pm 1}= -200$.
} \label{beta_c200}
 \vspace{5pt}
\end{figure}


For the $^{40}$K case studied in Ref.\cite{Ticknor04}, the
Feshbach resonances are at $B^*_0 \approx 198.8 {\rm G}$ and
$B^*_1 \approx 198.4 {\rm G}$.  There, $c_1$ and $c_0$ are both
only weakly field dependent and are approximately given by $-0.02
a_0^{-1}$ (see Fig \ref{mag}).
 Our choice of $\tilde c = - 100$ above corresponds to a
density of roughly $10^{-11} a_0^{-3}$ = $6.7 \times 10^{13} {\rm
cm}^{-3}$.
 The range of $-1/{\tilde v_0}$ for $\tilde \mu$ to go
from $0$ to $1$
corresponds to roughly $0.5$G
  if we take the same gas density.
(This field range is proportional to $n^{2/3}$ ).
 Near the resonant fields, $ -v_1^{-1} + v_0^{-1}
\approx 2.1 \times 10^{-8} a_0^{-3}$ and is roughly field
independent. Thus the density determines the values for both
$\tilde c_{0, 1}$ and $ -\tilde v_1^{-1} +\tilde v_0^{-1}$ while
varying the magnetic field corresponds roughly to moving along a
horizontal line on our phase diagram of Fig.\ref{beta_c} (with
increasing field towards the right and the distance of the line
from the horizontal axis proportional to $n^{-1}$).
For the density cited above, $ -\tilde v_1^{-1} +\tilde v_0^{-1}
\approx 2000$, hence we expect only the $k_z$ phase to be observed.
To observe the phase transition, we need a $^{40}K$ gas of
a much higher density, or another gas species with much smaller splitting
between the $m= 0$ and $m = {\pm 1}$ resonances.

In the above treatment, we have assumed that the terms with
explicit $1/|\alpha_m|^2$ factors in Eqs.(\ref{D0})-(\ref{n}) can
be dropped. The validity of this assumption has to
 be investigated by
first-principle calculations of two-atom collisions.
Following Ref \cite{chevy}, we
define the quantity
\begin{equation}
\eta_m \equiv \frac{M^2 |\alpha_m|^2}{\hbar^4 L^3}\sum_{\vec p}\frac{1}{p^2}
\   .
\end{equation}
$\eta_m$ quantifies the relative strength of coupling between the
$m$ closed-channel and the continuum.  With this notation, Eq.
(\ref{cm}) becomes
\begin{equation}
c_m = - \frac{4 \pi \hbar^4}{M^2 |\alpha_m|^2} (1 + \eta_m) \ .
\end{equation}
The single channel approximation used above corresponds to
$\eta_m \to \infty$. We now investigate the effect of finite $\eta_m$
on the many-body problem. Expressing the Eqs.(\ref{D0}),
(\ref{D1}), and (\ref{n}) by $\eta_m$ instead of $1/|\alpha_m|^2$,
we get
\begin{widetext}
\begin{eqnarray}
 \frac{M}{4\pi}D_0(-\frac{1}{v_0}+\frac{2Mc_0\mu}{\hbar^2})
 - \frac{\eta_0}{1+\eta_0}\frac{3M^3c_0}{40\pi^2\hbar^4}
 \left[D_0(2|D_1|^2+3|D_0|^2+2|D_{-1}|^2)-2D_{0}^*D_1 D_{-1}
   \right] \nonumber \\
 = \frac{2\pi}{L^3}\sum_{\vec k, m}D_m Y_{10}^*(\hat k) Y_{1
m}(\hat
 k) h(\vec k)
\end{eqnarray}
\begin{eqnarray}
 \frac{M}{4\pi}D_1(-\frac{1}{v_1}+\frac{2Mc_1\mu}{\hbar^2})
 - \frac{\eta_1}{1+\eta_1}\frac{3M^3c_1}{40\pi^2\hbar^4}
 \left[ D_1
(3|D_1|^2+2|D_0|^2+6|D_{-1}|^2)-D_{-1}^*D_0^2 \right] \nonumber \\
 = \frac{2\pi}{L^3}\sum_{\vec k, m}D_m Y_{11}^*(\hat k) Y_{1 m}(\hat
 k) h(\vec k)
\end{eqnarray}
\end{widetext}
\begin{eqnarray}
n &=& \frac{1}{2L^3} \sum_{\vec k} ( 1-\frac{\epsilon_k - \mu}{[
(\epsilon_k-\mu)^2+|\Delta_{\vec k}|^2]^\frac{1}{2}}
           - \frac{ |\Delta_{\vec k} |^2} { 2 \epsilon_k^2 } )
           \nonumber \\
  && - \frac{M^2}{(4\pi)^2\hbar^4}\sum_m
       \frac{2+\eta_m}{1+\eta_m} |D_m|^2 c_m
\label{neta}
\end{eqnarray}
We note here that the last term of Eq.(\ref{neta}) involves a
factor $(2 + \eta_m)/(1 + \eta_m)$.  This term actually arises
from two contributions.  The first one, involving $2/(1 +
\eta_m)$, is due to the closed-channel molecules (the last term in
Eq.(\ref{n})), and a second one,  involving $\eta_m /(1 + \eta_m)$
due to the open-channel atoms.

\vspace{50pt}
\begin{figure}[tbh]
\begin{center}
\includegraphics[width=3in]{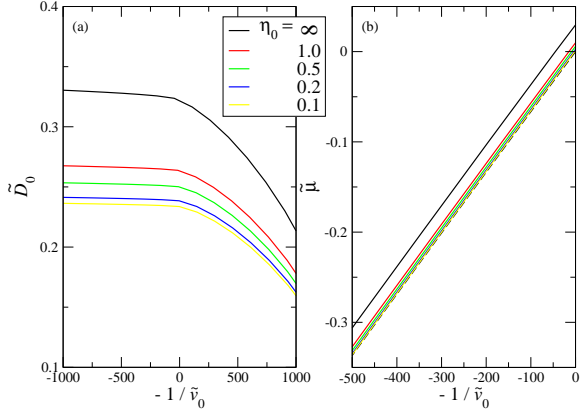}
\end{center}
 \caption{Plots of (a) $\tilde D_0$ and (b) $\tilde \mu$
 as functions of $-1/\tilde v_0$ at
 $\tilde c_0= \tilde c_{\pm 1}= -100$, $\tilde D_{\pm 1}=0$ for different
$\eta_0$'s.
$\eta_0=\infty$ corresponds that of Fig \ref{gap_c100}.
 The dashed line in (b) represents to the two-body result.}
\label{eta}
 \vspace{5pt}
\end{figure}


To show the effect of finite $\eta_0$, we plot the
general results for
 $\tilde D_0$ and $\tilde \mu$ in the state $\tilde
D_{\pm 1} = 0$  in Fig.\ref{eta}. The value of $\eta_0$ mainly
affects the results on the BEC side. In fact, deep in the BCS
limit the results are independent of $\eta_0$, as expected since
all atoms are basically in the open-channels. $\tilde \mu$ is
still basically linear in $ 1/ \tilde v_0$ with the same  slope,
and thus the width of the "crossover" region remains to be
determined by $\tilde c_0$ only (but not $\eta_0$). In the BEC
limit, when $\eta_0$ decreases from $\infty$, the contribution
from bound state molecules becomes more important (see in
particular Eq. (\ref{neta}) and the discussions below it).
 The chemical potential $\tilde
\mu$ becomes closer to the two-body value for decreasing $\eta_0$
as shown in Fig. \ref{eta}(b).
$\tilde D_0$ has the limiting value
$\simeq (32 \pi/3)^{1/3}
(- \tilde c_0)^{-1/2} [ (1 + \eta_0)/(2 + \eta_0) ]^{1/2}$,
and thus decreases with decreasing $\eta_0$.

The phase transition between the $k_z$ state and
the $k_z - i \beta k_y$ state can be investigated as before.
The value for $ -\tilde v_1^{* -1} +\tilde v_0^{-1}$
is not affected by the value of $\eta_m$'s in the
BCS limit.
 In the BEC limit,
if $\eta_0 = \eta_{\pm 1} \equiv \eta$,
we obtain $ -\tilde v_1^{* -1} +\tilde v_0^{-1}
\rightarrow (30.2) \eta/(2+\eta)$.
This critical value thus decreases with decreasing $\eta$.
Its behavior is as shown in Fig \ref{veta}.
The $\eta$ dependence of this transition line has also
been investigated in Ref \cite{gurarie}.

\vspace{70pt}
\begin{figure}[tbh]
\begin{center}
\includegraphics[width=3in]{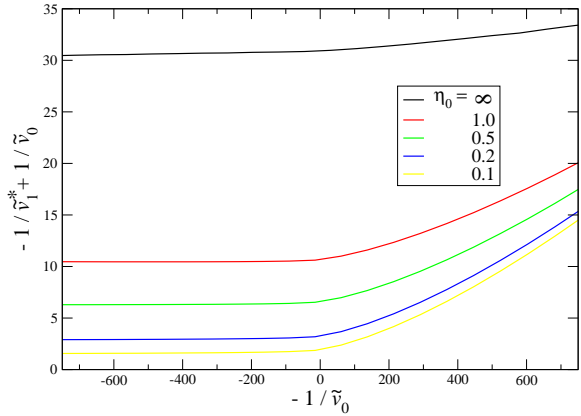}
\end{center}
 \caption{The transition line $ -\tilde v_1^{* -1} +\tilde v_0^{-1}$
between the $k_z$ phase and the $k_z - i \beta k_y$ phase
for $\eta$'s in the legend. $\tilde c_0 = \tilde c_{\pm 1} = -100$.}
\label{veta}
 \vspace{-5pt}
\end{figure}



We conclude with a discussion on the condition $\eta \gg 1$, or
equivalently $\hbar^4/(\alpha_m M)^2 \ll |c_m| $, for the validity
of the effective single-channel approximation. It is worth noting
that the corresponding condition for the s-wave Feshbach resonance
superfluidity is quite different.  For this latter case, the
single-channel approximation near resonanace is valid if
\cite{renorm,diener04,mackie05,szymanska05,partridge05} the
corresponding parameter $\alpha_s$ in Eq.(\ref{Hal}) satisfies
$\hbar^4/(|\alpha_s|^2 M^2) \ll 1/n^{1/3}$. This case is often
also referred to the "wide resonance" regime. Though for both s
and p-wave resonances, the single-channel approximation would be
valid for sufficiently strong coupling (large $\alpha$) between
the closed and open channels, for s-wave the condition for a
single-channel approximation is always satisfied when the gas is
sufficiently dilute. In contrast, for p-wave the condition $\eta
\gg 1$ is independent of the density of the gas. This difference
can be understood by noting that the dimension of $\alpha_s$ is
$\rm [E][L]^{3/2}$, whereas that of $\alpha_m$ here is $\rm
[E][L]^{5/2}$. The effective range $r_s$ for s-wave Feshbach
resonance  \cite{renorm,diener04} is proportional to
$\hbar^4/(|\alpha_s|^2 M^2)$, and the additional term $\propto
\sum (1/p^2)$ in Eq. (\ref{cm}) does not arise.

For the s-wave case, the terms "single-channel approximation"
and "wide-resonance" have been used interchangably.
However, we should note that for our p-wave case,
the condition $\eta \gg 1$ is not necessarily
in conflict with the fact that the
 the peak in scattering cross-section discussed in
Fig.1 of Ref.\cite{Ticknor04} is narrow.  This p-wave scattering
cross-section is proportional to $|f_m|^2$, which is given by
\begin{eqnarray}
|f_m|^2 = \frac{k^4}{(\frac{1}{v_m} + |c_m| k^2)^2 + k^6}
\end{eqnarray}
For $B > B_m^*$, where $1/v_m < 0$, the cross-section thus has a
peak at finite energy $E = \hbar^2 k_o^2/M$ where $k_o^2 = 1/ (|c_m v_m|)$.
The denominator
 can be written as $ c_m^2 (k^2 - k_o^2)^2 + k^6$.
For $B \gtrsim B_m^*$, $|f_m|^2$ as a function of $k^2$ then peaks
very near $k^2 = k_o^2$ and has a width approximately given by
$k_o^3/|c_m| = 1/ (|c_m|^{5/2}(-v_m)^{3/2})$ $\ll k_o^2$, and therefore a very
narrow (in energy) resonance. This is entirely a result of the
special form of $f$ for finite angular momentum.
   This narrowness of the peak is therefore not directly
connected with the assumption on the magitude of $\eta$ made
here.

In addition, the condition for $\eta \gg 1$ here is also {\em not}
necessarily in conflict with the assumption of a ``narrow" resonance in the
context discussed by Gurarie {\it et al} \cite{gurarie}.  They defined a
parameter $\gamma \propto \alpha_m^2 M^2 n^{1/3}$, and called the
case $\gamma \ll 1$ to be a narrow resonance.  Their condition is
fulfilled always at sufficiently low densities. Thus the resonance
can be narrow in their language whereas the single-channel
approximation can still be valid if $n^{1/3}  \ll \hbar^4/(\alpha_m M)^2
\ll |c_m| $.

\section{V. experimental probe of pairing symmetry }

It was suggested by Altman {\it et al.} \cite{altman} that
the measurement of the atom shot noise in the time-of-flight (TOF)
absorption image can reveal the properties of the many-body
states, including the symmetry of the pairing state.
 Later the proposal was carried out by Greiner {\it et al.}
\cite{greiner} for $^{40}$K atoms under an $s$-wave
Feshbach resonance. As discussed in previous section, the
pairing state induced by $p$-wave Feshbach resonance is either
$k_z$ or $k_z - i \beta k_y$ where $\beta$ can vary under a change
in the magnetic field or the gas density. In this section, we
propose how $\beta$ can be determined in a TOF experiment.

In the TOF experiment, one can measure the quantity
\begin{eqnarray}
{\cal G}(\vec r, -\vec r) &\equiv& \langle n(\vec r)n(-\vec
r)\rangle_t - \langle n(\vec r) \rangle_t \langle n(-\vec r)
\rangle_t \nonumber
\\ &\propto& \langle n_{\vec k}n_{-\vec k}\rangle -
\langle n_{\vec k} \rangle \langle n_{-\vec k} \rangle
\end{eqnarray}
where $\vec k \equiv m \vec r/(\hbar t)$. $t$ is the time after
the trapping potential and the interaction between atoms are
suddenly turned off. Using the mean-field
Hamiltonian in Eq.(\ref{hmf}), we have
\begin{eqnarray}
{\cal G}(\vec r, -\vec r) &\propto& \frac{|\Delta_{\vec
k}|^2}{(\epsilon_k - \mu)^2 + |\Delta_{\vec k}|^2} \nonumber \\
&=& \frac{\frac{3}{\pi}\tilde D_0^2 (\tilde k_z^2 + \beta^2 \tilde
k_x^2)}{(\tilde k^2 - \tilde \mu)^2 + \frac{3}{\pi}\tilde D_0^2
(\tilde k_z^2 + \beta^2 \tilde k_x^2)}
\end{eqnarray}
where $\tilde {\vec k} \equiv \vec k/k_{\rm F}$.
The two phases $k_z$ and $k_z - i \beta (\vec k \cdot \hat a)$
can be distinguished by the presence of anisotropy in
the $x-y$ plane.  For example, consider the weighted average
over $\phi$ defined by
\begin{equation}
S_0 (\alpha) \equiv \langle
{\rm cos} \left( 2 (\phi + \alpha) \right)
{\cal G}(\vec r, -\vec r) \rangle
\end{equation}
This average vanished in the $k_z$ phase.
For $\beta \ne 0$, $S_0 (\alpha)$ becomes finite
and can in principle be evaluated.
The presence of the $\tilde k_z^2 + \beta^2 \tilde k_x^2$
term in the denominator
complicates the analysis.  There are nevertheless
two possible simplifications.  In the BEC limit,
the terms involving $\tilde D_0^2$ in the denominator
are negligible compared with the first.
In the BCS limit we can confine ourselves to data where
 $r\gg R_{\rm F}$, equivalently $k\gg
k_{\rm F}$. In both cases the correlation function has
the angular dependence
\begin{eqnarray}
{\cal G}(\vec r, -\vec r)
&\propto&
(\tilde k_z^2 + \beta^2 \tilde k_x^2)
\end{eqnarray}
Then, for $\hat a = ({\rm cos} \chi) \hat y + ({\rm sin} \chi) \hat x$
as before, we get
\begin{equation}
S_0(\alpha) \propto - D_0^2 \beta^2 {\rm cos} \left( 2 (\chi - \alpha) \right)
\ .
\end{equation}
Thus $S_0(\alpha)$ will reveal the finiteness of $\beta$ as well
as the value of $\chi$ hence the direction of $\hat a$
(up to a sign).  The values of $\beta$ however are still difficult
to determine in this way, both because the need to evaluate
the proportionality coefficient in $S_0$ as well as the
fact that experiments so far detect a correlation in s-wave
substantially less than the expected theoretical value.
However, we can consider instead
\begin{eqnarray}
S_1 &\equiv& \langle \cos^2(\hat r\cdot
\hat z) {\cal G}(\vec r,-\vec r)  \rangle
 \\
S_2 &\equiv& \langle \sin^2(\hat r\cdot \hat z) {\cal G}(\vec
r,-\vec r) \rangle
\end{eqnarray}
These give
\begin{eqnarray}
\frac{S_1}{S_2}= \frac{3+\beta^2}{2+4\beta^2}
\end{eqnarray}
and hence the value of $\beta$ can be determined via
\begin{eqnarray}
\beta^2 = {\frac{3-2(S_1/S_2)}{4 (S_1/S_2) - 1}}
\end{eqnarray}
For the polar state, $\beta = 0$ and
$S_1 = 3 S_2/2 $.  For the axial state,  $\beta=1$ with $S_1=2S_2/3$.

\section{VI. Conclusion}

In conclusion, due to the magnetic dipolar interaction between two
fermionic atoms, the superfluidity in trapped gases induced by a
$p$-wave Feshbach resonance is expected to behave very differently
from the more familiar example of $^3$He. The symmetry of the
ground state can be tuned by varying either the density of the gas
or the applied external magnetic field. The phase can be a polar
state or a state intermediate between the polar and the axial
phase. We also propose how the pairing symmetry can be detected
from an experimental investigation.

\section{Acknowledgement}

S.K.Y. is very grateful to T.-L. Ho and A. J. Leggett for
useful discussions.
This research is supported by the National Science Foundation
under Grant No. PHY99-07949 and the National Science Council of
Taiwan under Grant Nos. NSC93-2112-M-001-016 and NSC94-2112-M-001-002.
The postdoctoral position of C.H.C. is supported
by NSC under Grant No.
NSC93-2816-M-001-0007-6 and Academia Sinica.



\end{document}